\documentclass[aps,prb,twocolumn,showpacs,amsmath,amssymb]{revtex4}
\usepackage{epsfig}
\usepackage{bm}
\def\be{\begin{equation}}
\def\ee{\end{equation}}
\def\bea{\begin{eqnarray}}
\def\eea{\end{eqnarray}}

\def\a{\alpha}
\def\b{\beta}
\def\g{\gamma}
\def\G {\Gamma}
\def\d{\delta}

\def \e{\varepsilon}
\def\h{}
\def\ph{\varphi}
\def\Tr{\mbox{tr}}
\def\S{{\cal S}}
\def\l{\lambda}
\def\L{\openone}
\def\m{\mu}
\def\o{\omega}
\def\O{\Omega}

\def\over {\overline}

\def\Bolts {}

\def\dwell{\tau_{\rm d}}

\begin{document}
\title{Shot noise of photon-excited electron-hole pairs in open quantum dots}
\date{\today}
\author{M. L. Polianski$^1$}
\email{polian@physics.unige.ch}
\author{P. Samuelsson$^2$}
\author{M. B\"uttiker$^1$}
\affiliation{$^1$D\'epartement de Physique Th\'eorique, Universit\'e de
Gen\`eve, CH-1211 Gen\`eve 4, Switzerland \\
$^2$Division of Solid State Theory, Lund University, S\"olvegatan 14
A, S-223 62 Lund, Sweden}

\begin{abstract}
We investigate shot noise of photon-excited electron-hole pairs in
open multi-terminal, multi-channel chaotic dots. Coulomb
interactions in the dot are treated self-consistently giving a
gauge-invariant expression for the finite frequency correlations.
The Coulomb interactions decrease the noise, the strong interaction
limit coincides with the non-interacting adiabatic limit. Inelastic
scattering and dephasing in the dot are described by voltage and
dephasing probe models respectively. We find that dephasing leaves
the noise invariant, but inelastic scattering decreases correlations
eventually down to zero.
\end{abstract}

\maketitle Investigation of noise\cite{review,Kozhev_Normal,glattli}
generally provides information not available from current
measurements, such as effective charge and quantum statistical
properties of the carriers. Most of the work has focused on noise in
the presence of stationary (dc) applied voltages. Of interest here
are shot noise measurements in photon-assisted
transport.\cite{review2} Initial experiments investigated noise in
the presence of both ac- and dc-potentials.\cite{Kozhev_Normal}
However, recently Reydellet \emph{et al.}\cite{glattli} reported
shot noise experiments in the presence of ac-potentials only. There
is no dc-current linear in voltage. Reydellet \emph{et al.}
\cite{glattli} subject one of the contacts of a two-terminal
conductor, a quantum point contact, to rf-radiation. Good agreement
was found between experiment and theory. \cite{LL,pedersen}

The purpose of our work is to investigate a {\it generic} conductor
with one or several contacts subject to ac-potentials and to
investigate the effect of Coulomb interactions, dephasing and
inelastic scattering on the photon-assisted noise. We consider a
multi-terminal chaotic quantum dot \cite{Beenakker, ABG} connected
to electronic reservoirs via quantum point contacts
$\alpha=1,2,..,M$ with a large number of channels (see Fig. 1). The
reservoirs are subject to oscillating potentials
$V_{\alpha}(t)=V_{\alpha}\cos(\omega t+\phi_{\alpha})$ at the same
frequency $\omega$, but with arbitrary phase $\phi_{\alpha}$ (this
is in contrast to the widely investigated case of shape modulating
potentials applied to gates \cite{Vavi}). In the absence of a
dc-bias, the elementary excitations generated by ac-potentials can
be understood as electron-hole ($e-h$) pairs rather than single
electrons.\cite{Mosk} The interpretation of the noise in terms of
$e-h$ pairs generated in the contacts of the sample provides an
intuitively appealing picture \cite{RPB} of the resulting
shot-noise. In the many-channel regime the noise correlator
$S_{\l\m}$ between the contacts $\lambda$ and $\mu$ fluctuates
weakly from sample to sample, and we consider only the
mesoscopically averaged shot-noise ${\over S}_{\lambda\mu}$.
\begin{figure}[b]
\centerline{\psfig{figure=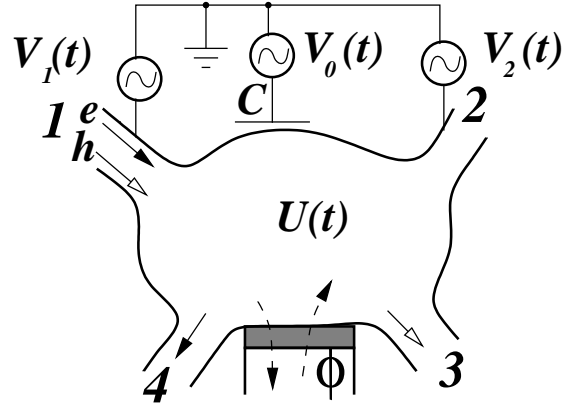,height=6cm}}
\caption{A four-terminal ($M=4$) chaotic dot is subject to
oscillating potentials $V_\a(t)$ at contacts $\a=1,2$ and coupled to
a gate with a time-dependent potential $V_0(t)$, via a capacitance
$C$. The internal potential of the dot is $U(t)$. Inelastic
scattering/dephasing/ with a rate $\g_\ph$ is introduced by
connecting a probe $\varphi$ to the dot. An electron-hole pair
excited in lead $1$ and split into leads $3$ and $4$ is indicated.}.
\label{fig2}
\end{figure}

The effect of inelastic scattering and/or dephasing on
photon-assisted noise has, to the best of our knowledge, not been
addressed in the literature. For dc-biased chaotic cavities, it was
shown that the ensemble averaged noise is insensitive to dephasing.
\cite{VLB,BS} It was expected \cite{BSB} and demonstrated
\cite{PSN,PJSB} that this is true not only for the conductance and
shot noise but also for higher order cumulants. To investigate
inelastic scattering and dephasing we employ here two models. In the
first, inelastic scattering is introduced by connecting the cavity
to a voltage probe.\cite{probe} The low frequency charge current
into the probe is zero. The voltage probe is characterized by a
(dimensionless) rate $\g_\ph=h/(\tau_\ph\Delta)$, where $\tau_\ph$
is the inelastic scattering time and $\Delta$ the mean level spacing
in the dot. In the second, dephasing probe model,\cite{dJB} in
addition to the charge current, the energy exchange with the probe
is prohibited. The rate $\g_\ph$ is then due to pure dephasing. In
both models the scattering is spatially uniform, which is ensured by
a large number $N_\ph\to\infty$ of poorly transmitting channels,
$\G\ll 1$, in the lead to the probe, such that $\g_\ph=N_\ph \G$ is
a finite constant.\cite{PietBeenakker} We compare the results from
random matrix theory with a semi-classical approach and find that
they lead to the same result.

It is convenient to express the noise in terms of an effective noise
temperature $T^*$. For the correlator ${\over S}_{\lambda\mu}$
between current fluctuations in the $\l$th and $\m$th leads, with
$N_\l$ and $N_\m$ channels respectively, we find for low frequencies
$\hbar\o\ll N\Delta$, with $N=\sum_\alpha N_{\alpha}$, and low
temperatures, to leading order in $eV_{\alpha}/\hbar \omega$
\begin{eqnarray}\label{eq:noise_deph}
{\over S}_{\l\m}^{\mbox{ }\rm deph}&=&2{\over G}_{\lambda\mu}k_B
T_0^*, \hspace{0.5 cm} {\over S}_{\l\m}^{\mbox{ }\rm volt}=2{\over
G}_{\lambda\mu}\frac{k_B T_0^* N}{N+\gamma_{\varphi}}, \\ \nonumber
T_0^*&=&\frac{e^2}{k_B\hbar\o}\left(\frac{\Tr \ V^2}{N}-\frac{|\Tr \
V e^{i\phi}|^2}{N^2}\right),
\end{eqnarray}
where $T_0^*$ is the noise temperature corresponding to the coherent
limit and ${\over G}_{\l\m}=(e^2/h)(N_\l\d_{\l\m}-N_\l N_\m/N)$, the
ensemble averaged differential conductance. Here we introduce
diagonal matrices of the amplitudes $V={\rm
diag}(V_1\L_1,...,V_M\L_M)$ and phase shifts $\phi={\rm diag
}(\phi_1\L_1,...,\phi_M\L_M)$, with $\L_\l$ a projector matrix on
the $\l$th lead.  Energy conserving dephasing processes do not
affect the noise to leading order in $N$, in particular the
dependence on the phases of the applied voltages is preserved. The
results from the dephasing probe model and the semi-classical
approach coincide, which demonstrates the consistency of the
different approaches and in particular supports the dephasing probe
model.\cite{note_probe}

 In contrast,
inelastic scattering suppresses noise similarly to the suppression
in conductors subject to a dc voltage only.\cite{review} The
decrease of the noise temperature with increasing inelastic rate can
be understood as follows: a sufficiently large voltage probe acts as
an absorber of rare $e-h$ pairs created in the real leads. An
absorption of these pairs creates no potential response. Only events
in which a single electron or hole reaches the voltage probe give
rise to potential fluctuations and to the generation of $e-h$ pairs
by the probe. With increasing coupling to the probe it is ever more
likely that both the electron and the hole generated in a real lead
end up in the voltage probe, rather then split into
current-measuring leads. For the dephasing probe the balance of the
energy currents is reached through a non-equilibrium distribution in
the probe. The probe serves as a source of $e-h$ pairs which can be
split into different outgoing leads, giving an additional
contribution to the correlations which exactly compensates for the
absorbed pairs.

AC-voltages can generate excess charge densities and, therefore, it
is important to treat the effect of Coulomb interactions on the
noise. Experiments can be performed at sufficiently high frequencies
for which the conductance is energy-dependent. Then a
non-interacting treatment gives an unphysical (not gauge-invariant)
result. To have a meaningful result, also including possible
external macroscopic gates, we consider Coulomb interactions on the
level of the random phase approximation \cite{Hartree} and find the
dynamic self-consistent potential $U(t)$ in the sample. For open,
many channel conductors, the neglected part of the Coulomb
interactions (Fock-terms) gives only sub-leading corrections
\cite{BLF} and is disregarded here.

We find a signature of Coulomb interactions when the period of the
ac-excitation is comparable to the charge relaxation time $\tau$,
rather than the dwell time $\dwell$ of the quantum dot. A
qualitatively similar result was found for the ac-conductance in a
quantum dot.\cite{PietMarkus} The interactions decrease the
correlations: In the absence of interactions the current
fluctuations in different leads are uncorrelated, since there is no
requirement of charge conservation in the dot. With interactions,
these fluctuations are suppressed by negative feedback from
displacement currents due to the internal potential. We remark that
for strong Coulomb interaction, $\tau\to 0$, the noise corresponds
effectively to the low frequency limit $\hbar\o\ll N\Delta$ for the
non-interacting system. This explains why the expression for the
noise in the ac-driven system of Refs.  [\onlinecite{LL,pedersen}]
fits the experimental data [\onlinecite{Kozhev_Normal,glattli}] so
well: the dispersion of the noise is significant on a much larger
frequency scale than applied in these experiments.

 \emph{Formalism} We
 first consider the current correlations in the leads $\l,\m$ using
 scattering theory.\cite{review,Beenakker} The coherent open dot is
 fully characterized by its unitary $N\times N$ scattering matrix
 $\cal S$. The dot is in the chaotic regime, \emph{i.e.} the dwell time of
 the dot $\dwell=h/(N\Delta)$ is sufficiently large. Scattering is
 spin-independent and the results given below are presented for a single spin
 direction. We take $e=h=k_{\rm B}=1$ and express the
 sample-specific noise in terms of energy-dependent scattering
 matrices ${\cal S}(\e)$, amplitudes of applied voltages $V_\a$, their
 relative phase shifts $\phi_\a$ and electronic distributions in the
 leads $f_\a(\e)$:\cite{RPB}
\begin{eqnarray}\label{eq:ij}
S_{\lambda\mu}&=&\sum_{klm,\,\alpha\beta}\int d\e \,\Tr\left({\bf
A}_{\alpha\beta}(\lambda,\e-m\h\o){\bf
A}_{\beta\alpha}(\mu,\e)\right)\nonumber \\
&\times&
J_{k}\left(\frac{V_\a}{\h\o}\right)J_{k+m}\left(\frac{V_\a}{\h\o}\right)
J_{l+m}\left(\frac{V_\b}{\h\o}\right)J_{l}\left(\frac{V_\b}{\h\o}\right)\nonumber\\
&\times & e^{im(\phi_\b-\phi_\a)}f_\a(\e-k\h\omega)(1- f_\b
(\e-l\h\omega)),
\end{eqnarray}
where ${\bf A}(\l,\e)=\L_\l-\S^\dag(\e) \L_\l \S(\e)$ and $J_n$ is
the $n$th order Bessel function.  The noise power Eq. (\ref{eq:ij})
differs from sample to sample according to its mesoscopic
distribution $P(S_{\l\m})$. However, for $N\gg 1$ the value of the
ensemble averaged correlations ${\over S}_{\l\m}$ is
representative\cite{note} (the average is carried following Refs.
[\onlinecite{diagram,iop}]). We note that to leading order in $N\gg
1$ the averaged noise is unaffected by a time-reversal symmetry
breaking magnetic field. Below we first consider the effect of
Coulomb interaction and later find the role of decoherence
(dephasing) on the averaged noise ${\over S}_{\lambda\mu}$. We also
show that the noise can be found within a semi-classical approach.

\emph{Coulomb interactions} First we find the correlations in the
non-interacting limit, when the internal potential of the dot and
its coupling to external gates are not accounted for. We obtain the
effective noise temperature $T_{\rm NI}^*$, with ${\over
S}_{\lambda\mu}=2{\over G}_{\lambda\mu} T^*_{\rm NI}$, as
\begin{eqnarray}\label{eq:Main}
\Bolts T_{\rm NI}^*&=& \frac{1}{\h\o}\left(\frac{\Tr \ V^2}{N}-\frac
{|\Tr\ V e^{i\phi}|^2}{N^2+\o^2/\Delta^2}\right).
\end{eqnarray}
In the limit $\omega\ll N\Delta$ the noise temperature $T_{\rm
NI}^*\rightarrow T_0^*$, with $T_0^*$ defined in Eq.
(\ref{eq:noise_deph}). Only in this limit is the noise
gauge-invariant, \emph{i.e.} unaffected by a uniform shift of
applied potentials. At finite frequencies it is necessary to
consider the internal potential $U(t)$ to obtain a physically
meaningful result. The interacting problem with $U(t)\neq 0$ is
reduced to the noninteracting one by the global uniform shift
$V(t)\to V(t)-U(t)$ in Eq. (\ref{eq:ij}). The potential $U(t)$ is
found using gauge invariance of currents. Charge inflow into the dot
shifts the potential $U(t)$ due to capacitive coupling to the gate,
kept at a potential $V_0(t)=V_0\cos(\o t+\phi_0)$. The current
$I_\a(\O)$ at finite frequency $\O$ is the sum of particle currents
$\sum_\b G_{\a\b}(\O)V_\b(\O)$ and the displacement current
$\chi_\a(\O) U (\O)$ due to variations of the uniform potential in
the dot, with \cite{Hartree}
\begin{eqnarray}\label{eq:G}
G_{\a\b}(\O)= N_\a \d_{\a\b}-\int_{0}^{\h\O}\frac {d\e}{\h\O} \Tr
\,\openone_\a{\cal S}^\dagger(\e-\h\O)\openone_\b{\cal S}(\e).
\end{eqnarray}
Gauge invariance implies a susceptibility given by \cite{Hartree}
$\chi_\a(\O)=-\sum_{\beta}G_{\alpha\beta}(\Omega)$, and the current
$I_\a$ reads
\begin{eqnarray}
I_\a(\O)&=& \sum_\b G_{\a\b}(\O)(V_\b(\O)-U(\O)).
\label{eq:invar_I}
\end{eqnarray}
We note that to leading order in $V/\omega$ only $\Omega=\pm \o$
contribute. From charge conservation
$\sum_{\alpha}I_{\alpha}(t)=C(d/dt)[V_0(t)-U(t)]$, we find the
sample-specific potential $U_\o$. The potential $U_\o$ is
self-averaging, \emph{i.e.} its sample-to-sample fluctuations can be
neglected, if $N^2|\Tr (Ve^{i\phi})|^2\gg\Tr\ ( V^2e^{2i\phi})$ and
$N\gg 1$. Here we assume that this is the case and hence the
potential $U_\o$ averages to
\begin{eqnarray}\label{eq:Uomega}
{\over U}_\o&=& V_0 e^{i\phi_0}+\frac{C_\mu/C }{1-i\o R_q
C_\mu}\left(\frac{\Tr\,V e^{i\phi}}{N}-V_0 e^{i\phi_0}\right).
\end{eqnarray}
We have introduced the electrochemical capacitance \cite{Hartree}
$1/C_{\mu} = \Delta +1/C$, the series addition of the geometrical
capacitance $C$ and the "quantum capacitance" determined by density
of states of the cavity $1/\Delta$, as well as the charge relaxation
resistance \cite{Hartree,PietMarkus}$R_q=1/N$. The Coulomb
interaction thus leads to charging processes on the scale of the
charge relaxation time $\tau=R_q C_{\mu}$. Using now this Hartree
potential Eq. (\ref{eq:Uomega}) we find that the Coulomb interaction
modifies the non-interacting result (\ref{eq:Main}), in that the
density of states $1/\Delta$ is substituted \cite{NPB} by the
electrochemical capacitance $C_\m$. Taking into account the external
gate potentials then yields a gauge-invariant noise temperature
\begin{eqnarray}\label{eq:Sint}
 \Bolts T_{\rm I}^*&=&
T_0^*+\frac{1}{\o}\frac {|\Tr(V e^{i\phi})/N-V_0 e^{i\phi_0}|^2 (\o
R_q C_\m )^2}{1+(\o R_q C_\m )^2}.
\end{eqnarray}
An important test of validity of Eq.  (\ref{eq:Sint}) is obtained by
considering the limit of synchronous voltages at all contacts,
$V_{\alpha}=V, \phi_{\alpha}=0$. This corresponds to a global shift
of the potential and has no physical consequences.

Since typically $C_\m\Delta\ll 1$, this implies that much higher
frequencies are needed to observe the dispersion of the shot noise
than one might naively expect from the non-interacting result
(\ref{eq:Main}). As a consequence, for the experimentally relevant
limit $\o\tau\ll 1$, the noise temperature $T_{\rm I}^*$ coincides
with the result $T_0^*$ for the adiabatic cycling of the potentials,
$\omega \ll N\Delta$, in the non-interacting limit. Below we focus
on this low frequency limit $\omega \tau \ll 1$ when deriving the
result in Eq. (\ref{eq:noise_deph}) for dephasing and inelastic
scattering. As described above, the dephasing and inelastic
scattering is modeled by connecting the dot to voltage and dephasing
probes. Both models require that the low frequency charge current
into the additional lead $\ph$ vanishes. For a real voltage probe
this corresponds to a voltmeter with infinite impedance at zero
frequency, dropping off at higher frequencies. If there is no
parallel capacitance to the probe, then the potentials $V_{\varphi}$
of the probe and the dot must be equal and are obtained from Eq.
(\ref{eq:Uomega}) in the limit $V_0=0,\,C\Delta \to 0,\,\o\to 0$.

\emph{Voltage probe model} Scattering in the quantum dot gives rise
to bare particle current fluctuations in the leads, $\delta
I_{\alpha}$. At low frequencies, the potential $\delta V_{\varphi}$
in the voltage probe fluctuates to maintain zero current. The {\it
real } leads are however voltage controlled and $\d V_\alpha=0$.
Thus the total fluctuation $\Delta I_{\alpha}(t)$ of the current in
the real leads consists of particle current fluctuations $\delta
I_\alpha$ and additional displacement current fluctuations
$G_{\a\ph}\d V_\ph$ [Note that $G_{\a\ph}$ is the conductance for
the dot connected via a non ideal contact to the probe]. The
conservation of current fluctuations into the probe, $\Delta
I_{\varphi}=0$, gives $\sum_\a G_{\ph\a}\d V_\ph(t)=-\d I_\ph(t)$
from which $\delta V_{\varphi}$ is determined. We can then write the
total fluctuations $\Delta\vec I(t)={\tilde \L} \d\vec I(t)$ with $
{\tilde \L_\l}=\L_\l+ \L_\ph G_{\l\ph}/G_{\ph\ph}$, where we
introduced a vector notation $\Delta\vec I(t)=[\Delta
I_1(t),...,\Delta I_M(t)]$. Using Eq. (\ref{eq:ij}) in the low
frequency limit, the total noise $S_{\lambda\mu}^{\rm volt}=\langle
\Delta I_{\lambda}\Delta I_{\mu}\rangle$ is given by
\begin{eqnarray}\label{eq:NNoise_deph}
 S_{\l\m}^{\rm volt}
 &=&\sum_{\a\b}
\frac{|V_\a e^{i\phi_\a}-V_\b e^{i\phi_\b}|^2}{4\o} 
\Tr({\tilde{\bf A}(\l)}_{\a\b}{\tilde{\bf A}(\m)}_{\b\a}),
\end{eqnarray}
where ${\tilde{\bf A}(\l)}={\tilde \L_\l}-{\cal S}^\dagger{\tilde
\L_\l}{\cal S}$. For a coherent dot, the effective noise temperature
in the limit $\o\ll N\Delta$ is given by $T_0^*$, while for a
partially coherent dot the effective temperature is suppressed,
given by $NT_0^*/(N+\gamma_{\varphi})$, as stated in the r.h.s. of
Eq. (\ref{eq:noise_deph}). This result is obtained from separate
averaging of the matrices $\tilde\L$ and the 4-matrix correlators
for a quantum dot with non-ideal leads.\cite{notyet}

\emph{Dephasing probe model} We now compare the results for the
inelastic voltage probe with the dephasing probe model.
Current conservation at each energy now determines the
non-equilibrium distribution function $f_{\varphi}(\e)$ in the
dephasing probe. Mesoscopic fluctuations of $f_\ph$ are small, so
that we can characterize a dot by the mesoscopically averaged
distribution $\over f_\ph(\e)$. The detailed balance of the currents
at energy $\e$ leads to
\begin{eqnarray}\label{eq:f}
\sum_m J_m^2\left(\frac{V_\ph}{\o}\right){\over f_\varphi(\e-m\o)} =
\sum_{\b,\,m>\,\e/\o
}\frac{N_\b}{N}J_m^2\left(\frac{V_\b}{\o}\right),
\end{eqnarray}
where the summation is taken over real leads $\b$ only. To leading
order in $V/\omega$ the distribution is
\begin{eqnarray}\label{eq:f_aver}
\over f_\varphi(\e)=
\Theta (-\e)+\mbox{sgn }\e\ \Theta(\omega-|\e|)\frac{\Bolts
T_0^*}{\o}.
\end{eqnarray}
The fluctuations of the distribution function are treated in a
similar way as for the voltage probe. However, the correlations in
the dephasing probe model have both the contribution
(\ref{eq:NNoise_deph}) due to the potential $\overline V_\o$ and a
term due to the non-equilibrium distribution $\overline f_\ph$. As a
consequence, the noise is characterized by $T_0^*$, \emph{i.e.}
elastic dephasing does not affect the noise.

We note that in an experiment, a finite temperature $T$ affects the
noise measurements. From Eq. (\ref{eq:ij}) taken at finite
temperature $T$ the equilibrium noise temperature is readily obtained,
$T_{\rm eq}^*=T$. Temperature effects can thus be neglected for
inelastic scattering if $T\ll NT_0^*/(N+\gamma_{\varphi})$ and for
$T\ll T_0^*$ in the absence of inelastic scattering.


\emph{Semi-classical approach} The fact that $N\gg 1$ and the
observation that the noise is insensitive to elastic dephasing
suggests that, just as in the dc-biased case,\cite{BS,BSB,PSN,PJSB}
a semi-classical description of the noise is possible. This is
indeed the case, when $V^2/\o\gg \Delta$.\cite{RPB} The point
contacts act as independent emitters of fast fluctuations into the
dot. The distribution function $f(\e,t)=f(\e)+\delta f(\e,t)$ of the
dot responds with fluctuations $\delta f(\e,t)$ to preserve current
at each energy. The total charge current into the probe is zero,
which gives the potential of the dot
$U(t)=\Tr\,V(t)/N$. The static part of the 
distribution function $f(\e)$ coincides with the distribution
function $\over f_{\varphi}(\e)$ in Eq. (\ref{eq:f_aver})
, since $\over f_{\varphi}(\e)$ was derived under the assumption
that currents are conserved at each energy.

The total fluctuations of the current $\Delta I_\a(\e)$ in lead $\a$
at energy $\e$ are the sum of bare fluctuations $\d I_\a(\e)$ and
the displacement current fluctuations $N_\a\d f(\e,t)$, $\Delta
I_\a(\e)=\d I_\a(\e)+N_\a\d f(\e,t)$. Conservation of current at
each energy in the dot gives $\d f(\e,t)=-(1/N)\sum_{\alpha}\d
I_\a(\e)$, yielding $\Delta I_\a(\e)$ in terms of the bare
correlators $\delta I_\a(\e)$. The total noise ${\over
S}_{\lambda\mu}=\langle \Delta I_{\lambda}\Delta I_{\mu}\rangle$ is
thus given by a weighted sum of the correlators of the individual
point contacts $\langle \delta I_{\alpha}\delta
I_{\beta}\rangle=\delta_{\alpha\beta}S_{\alpha}$, with $S_{\alpha}$
given by Eq. (\ref{eq:ij}) for the corresponding two terminal point
contact $\alpha$. Due to the non-equilibrium distribution function
$f(\e)$ in the dot, the noise $S_{\alpha}$ is nonzero. Summing up
all the terms in $\bar S_{\lambda\mu}$ we arrive at the effective
noise temperature $T_0^*$, the same as obtained within a random
matrix approach. Similarly we can show that a voltage probe
suppresses the correlations, giving a noise temperature
$NT_0^*/(N+\gamma_{\varphi})$ in accordance with Eq.
(\ref{eq:noise_deph}).

In conclusion, we have investigated the shot noise of $e-h$ pairs in
chaotic quantum dots subject to an ac-bias at the contacts. The
noise has been derived within a scattering random matrix approach as
well as with a semi-classical approach. By including Coulomb
interactions, a gauge invariant theory for finite frequencies was
constructed. It was found that elastic dephasing does not affect the
noise, however inelastic scattering leads to a suppression of the
noise.

We would like to thank Piet Brouwer and Eugene Sukhorukov for
important discussions, and Leo DiCarlo and Misha Moskalets for their
comments on the manuscript. This work was supported by the Swiss NSF
and the Swedish VR.


\begin{thebibliography}{MMM}
\bibitem{review} Ya.~M.~Blanter and M.~B\"uttiker, Phys.~Rep. \textbf{336}, 2
(2000).

\bibitem{Kozhev_Normal} R.~J.~Schoelkopf, A.~A.~Kozhevnikov, D.~E.~Prober, and M.~J.~Rooks, Phys.~Rev.~Lett. \textbf{80},
2437 (1998).

\bibitem{glattli} L.-H.~Reydellet, P.~Roche, D.~C.~Glattli,
B.~Etienne, and Y.~Jin, Phys.~Rev.~Lett. \textbf{90}, 176803 (2003).

\bibitem{review2} G.~Platero and R.~Aguado, Phys.~Rep. \textbf{395}, 1
(2004).

\bibitem{LL} G.~B.~Lesovik and L.~S.~Levitov, Phys.~Rev.~Lett.
\textbf{72}, 538 (1994).

\bibitem{pedersen}  M.~H.~Pedersen and M.~B\"uttiker, Phys.~Rev.~B
\textbf{58}, 12 993 (1998).



\bibitem{Beenakker} C.~W.~J.~Beenakker, Rev. Mod. Phys. {\bf 69}, 731 (1997).

\bibitem{ABG} I.~L.~Aleiner, P.~W.~Brouwer, and L.~I.~Glazman,
Phys.~Rep. {\bf 358}, 309 (2002).

\bibitem{Vavi}M.~G.~Vavilov, L.~DiCarlo, and C.~M.~Marcus,
Phys.~Rev.~B {\bf 71}, 241309(R) (2005).

\bibitem{Mosk} M. Moskalets and M. B\"uttiker, Phys. Rev. B {\bf 66}, 035306, (2002); {\it ibid} {\bf 70}, 245305 (2004).

\bibitem{RPB} V.~Rychkov, M.~L.~Polianski, and M.~B\"uttiker,
cond-mat/0507276.

\bibitem{VLB}S.~A.~van~Langen and M.~B\"uttiker, Phys.~Rev.~B
\textbf{56}, R1680 (1997).

\bibitem{BS} Ya.~M.~Blanter and E.~V.~Sukhorukov, Phys.~Rev.~Lett. {\bf 84}, 1280
(2000).
\bibitem{BSB}Ya.~M.~Blanter, H.~Schomerus, and C.~W.~J.~Beenakker, Physica~E {\bf 11}, 1
(2001).

\bibitem{PSN}K.~E.~Nagaev, P.~Samuelsson, and S.~Pilgram, Phys.~Rev.~B {\bf 66}, 195318
(2002).

\bibitem{PJSB}S.~Pilgram, A.~N.~Jordan, E.~V.~Sukhorukov, and M.~B\"uttiker,
Phys. Rev. Lett. {\bf 90}, 206801 (2003).

\bibitem{probe}M.~B\"uttiker, Phys. Rev. B {\bf 33}, 3020 (1986).

\bibitem{dJB} M.~J.~M. de Jong and C.~W.~J.~Beenakker, Physica A {\bf 230},
219 (1996).

\bibitem{PietBeenakker} P.~W.~Brouwer and C.~W.~J.~Beenakker, Phys.~Rev.~B {\bf 55}, 4695
(1997).

\bibitem{note_probe}For {\em few-channel} systems predictions of the dephasing probe model
might differ from phase averaged results, see F.~Marquardt and
C.~Bruder, Phys.~Rev.~B {\bf 70}, 125305 (2004). See also
C.~W~J.~Beenakker and B.~Michaelis, cond-mat/0503347.

\bibitem{Hartree} M.~B\"uttiker, A.~Pr\^etre, and H.~Thomas, Phys.
Rev. Lett. {\bf 70}, 4114 (1993).

\bibitem{BLF} P.~W.~Brouwer, A.~Lamacraft, and K.~Flensberg, Phys.~Rev.~B {\bf 72}, 075316 (2005).

\bibitem{PietMarkus} P.~W.~Brouwer and M.~B\"uttiker, Europhys.\ Lett. {\bf 37},
441 (1997).
\bibitem{note} Although formally considering $N_\a\gg 1$, we expect
the result to be useful for few-channel dots (see discussion of
conductance for $N=2$ in Refs. [\onlinecite{Vavi,BLF}]).

\bibitem{diagram} P.~W.~Brouwer, C.~W.~J.~Beenakker, J.~Math.~Phys. {\bf 37}, 4904 (1996).

\bibitem{iop} M.~L.~Polianski and P.~W.~Brouwer, J.~Phys.~A:~Math.~Gen. {\bf 36}, 3215
(2003).


\bibitem{NPB} Interestingly in the 3rd cumulant of the dc-biased cavity both $\dwell$ and
$\tau$ appear: K.~E.~Nagaev, S.~Pilgam, and M.~B\"uttiker,
Phys.~Rev.~Lett. {\bf 92}, 176804 (2004); S.~Pilgram, K.~E.~Nagaev,
and M.~B\"uttiker,  Phys.~Rev.~B {\bf 70}, 045304 (2004); B.~Reulet
and D.~ E.~Prober, Phys. Rev. Lett. {\bf 95}, 066602 (2005).


\bibitem{notyet} M.~B\"uttiker and M.~L.~Polianski, cond-mat/0508220.
\end{thebibliography}
\end{document}